\documentclass[12pt]{article}
\usepackage{amssymb}
\usepackage{epsfig}
\usepackage{graphicx}
\usepackage{amsmath}
\usepackage{verbatim}

\csname @addtoreset\endcsname{equation}{section}
\begin{document}
\begin{titlepage}
\title{\bf\Large  Phenomenologies of Higgs Messenger Models \vspace{18pt}}

\author{\normalsize Sibo~Zheng , Yao Yu and Xing-Gang Wu \vspace{12pt}\\
{\it\small  Department of Physics, Chongqing University, Chongqing 401331, P.R. China}\\}

\date{}
\maketitle \voffset -.3in \vskip 1.cm \centerline{\bf Abstract}
\vskip .3cm In this paper, we investigate the phenomenologies of
models where the Higgs sector plays the role of messengers in gauge
mediation. The minimal Higgs sector and its extension are considered
respectively. We find that there exist viable models when an
appropriate parity is imposed. Phenomenological features in these
kind of models include three sum rules for scalar
masses, light gluino as well as one-loop $\mu$ and two-loop $B\mu$
terms. \vskip 5.cm \noindent  01/ 2011
 \thispagestyle{empty}

\end{titlepage}
\newpage

\section{Introduction}
Among mediation mechanism of supersymmetry (SUSY-) breaking ,
gauge mediation \cite{SUSY1,SUSY2,SUSY3,SUSY4,SUSY5,SUSY6,SUSY7,
SUSY8,SUSY9,SUSY10} is an attractive scenario
that can naturally address the stringent flavor problem of TeV-scale new physics,
in addition solve the mass hierarchy between Plank and electroweak scale.
It can also realize the grand unification of gauge couplings under some circumstances.
However, the collider signals are quite dependent on the details of microscopic models,
despite there exists a systematic method for calculating the sfermion and gaugino masses
in this scenario \cite{GM1}.

Recently, in \cite{1012.2952} a specific model is proposed,
in which the Higgs sector $H_{\mu,d}$ together with the messenger sector mediate the SUSY-breaking effects.
One reason that Higgs fields assist gauge mediation is that the messengers under bi-fundamental $SU(5)$
representation probably mix with the Higgs fields via multiple SUSY-breaking $F$-terms.
In this model, the gaugino masses are the same as in ordinary gauge mediation while
the sfermion masses receive a new part of contribution through Yukawa interactions in superpotential of
minimal supersymmetric standard model (MSSM).
A distinct mass spectra can be expected in this model.
In this sense, it is appealing phenomenologically.

In this paper, we will proceed to investigate possible variations based on the idea
that the Higgs sector serves as messengers in gauge mediation.
A direct consideration is that there are no extra messengers except the Higgs sector,
which we refer to as minimal Higgs messenger models.
Since the contribution to sfermion masses coming from the Yukawa interactions is negative \cite{1012.2952}.
There is a competition between the contributions arising from gauge mediation and Yukawa interactions.
When taking the phenomenological constraints into account,
we find that there is no viable parameter space.
The most serious problem is how to reconcile the electroweak symmetry breaking (EWSB) with extremely large
$\mu$ and $B\mu$ terms.

Instead of solving the problem via introducing $SU(5)$ messengers and multiple $F$-terms,
we discuss another possibility of extending the Higgs sector.
Explicitly, we introduce two extra Higgs doublets $R_{\mu,d}$.
By imposing a discrete $Z_4$ symmetry,  we make sure that
$R_{\mu,d}$ doublets directly couple to the SUSY-breaking spurion superfield while the $H_{\mu,d}$ not.
In this setup, there are negligible flavor violations and distinctive phenomenological features in this model.
We find that $i)$, there are three sum rules for sfermion masses. $ii)$,
In contrary to one-loop generation of gaugino masses in ordinary gauge mediation,
the gluino mass $m_{\lambda_3}$ are generated at three-loop and the other two gauginos masses are generated at one-loop.
\footnote{See \cite{1101.1645} for recent discussion about stable standard model charged superparticles.}
$iii)$, the $\mu/B\mu$ terms can be correctly reproduced with the help of $Z_4$ parity.

The paper is organized as follows. In section 2, we discuss the minimal Higgs messengers.
The scalar and gaugino masses formulas are obtained and the parameter space is analyzed.
In section 3, first we discuss the setup of our model, then analyze the phenomenologies.
The last part of section 3 is devoted to reproduce the one-loop $\mu$ and two-loop $B\mu$ terms in
Higgs messenger models.

\section{Minimal Higgs messenger models}
First we consider the minimal Higgs messenger models, in which the
Higgs superfields directly couple to SUSY-breaking sector $X=M+\theta^{2}F$ via
tree-level superpotential,
\begin{eqnarray}{\label{e1}}
W=XH_{\mu}H_{d}
\end{eqnarray}
The scalars of Higgs superfields obtain masses
$\phi_{\pm}=M^{2}\pm~F$ under eigenvector $(H_{\mu}\pm
H_{d})/\sqrt{2}$, while Higgsinos obtain degenerate masses $M$.
The SUSY-breaking effects are mediated to visible sector via Higgs superfields.
In addition to generating  one-loop gaugino and two-loop sfermion masses through
the vector superfields of SM \cite{9607225, 9608224, DM},
there are extra one-loop contributions to sfermion terms arising from
Yukawa interactions through superpotential of MSSM  $W_{MSSM}$ \cite{1012.2952}.

Collect these two sets, we obtain the total contributions to sfermion masses,
\begin{eqnarray}{\label{e2}}
\tilde{m}^{2}_{Q_{i}}=M^{2}\left[\left(y^{2}\sum_{r=1,2}C^{r}(f_{i})\left(\frac{g_{r}^{2}}{16\pi^{2}}\right)^{2}\right)\times~f(y)+
\left(\frac{Y^{2}}{32\pi^{2}}\right)\times~g(y)\right]
\end{eqnarray}
with
\begin{eqnarray}{\label{e3}}
g(y)&=&(2+y)\ln(1+y)+(2-y)\ln(1-y)\nonumber\\
f(y)&=&\frac{1+y}{y^{2}}\left[\ln(1+y)-2Li_{2}\left(\frac{y}{1+y}\right)+\frac{1}{2}Li_{2}\left(\frac{2y}{1+y}\right)\right]+(y\rightarrow -y)
\end{eqnarray}
where $C^{r}$ is the quadratic Casimir invariant of MSSM and parameter $y=F/M^{2}$.
$Y$s are Yukawa coupling constants in superpotential $W_{MSSM}$.
The positivity of Higgs scalar masses requires that $y<1$.
Note that $g(y)$ is negative value in parameter space $0<y<1$.

The gaugino masses of $SU(2)\times~U(1)$ gauge group  are the same as before,
they are generated at one-loop,
\begin{eqnarray}{\label{e4}}
m_{\lambda_{r}}=\left(\frac{g_{r}^{2}}{16\pi^{2}}\right)M\times\left[\frac{1+y}{y}\ln(1+y)+(y\rightarrow -y)\right], (r=1,2)
\end{eqnarray}
In the case of minimal Higgs messenger models, where $H_{\mu,d}$ server as the messengers, there is no way out for EWSB in these models, as we will discuss hereafter,
we do not address the gluino mass in this scheme.
In the case of variant Higgs messenger models that will be explored in section 3,
the leading-order gluino mass is generated at
three-loop.\footnote{ The reason for this is due to fact that
there are no irreducible two-loop corrections to $g_{3}$ with messengers as internal lines in
two-loop Feynman diagrams, also there are
no renormalizable Yukawa couplings for messengers as for Higgs doublets $H_{\mu,d}$.
}
Now we derive the gluino mass in terms of the wave-function renormalization \cite{GR}.
The Renormalization group equation of QCD gauge coupling up to three-loop is given by,
\begin{eqnarray}{\label{A1}}
\frac{dg_{3}}{dt}=\frac{g^{3}_{3}}{16\pi^{2}}B^{(1)}+\frac{g^{3}_{3}}{(16\pi^{2})^{2}}\sum_{r=1,2,3}B^{(2)}_{r}g_{r}^{2}
+\frac{g^{3}_{3}}{(16\pi^{2})^{3}}\sum_{r=1,2}B^{(3)}_{r}g^{4}_{r}+\cdots
\end{eqnarray}
where we neglect other three-loop contributions that are irrelevant for present discussion about gluino mass.
The value of $B^{(1)}$ and  $B^{(2)}_{r}$ is explicitly given in \cite{Martin93}.
The coefficients $B^{(3)}_{r}$ contain the correction arising from messenger fermionic and bosonic loops.
It follows from eq.\eqref{A1},
\begin{eqnarray}{\label{g1}}
\frac{d \alpha_{3}^{-1}}{dt}=-\frac{1}{2\pi}B^{(1)}-\frac{1}{2\pi(16\pi^{2})}\sum_{r=1,2,3}B^{(2)}_{r}g_{r}^{2}
-\frac{1}{2\pi(16\pi^{2})^{2}}\sum_{r=1,2}B^{(3)}_{r}g^{4}_{r}+\cdots
\end{eqnarray}
Solving eq.\eqref{g1} one obtains,
\begin{eqnarray}{\label{g2}}
16\pi~S(X,\mu)&=&\frac{\alpha^{-1}(\Lambda_{UV})}{16\pi}-\frac{1}{2\pi(16\pi^{2})^{2}}
\sum_{r=1,2}B^{(3)}_{r}g_{r}^{4}\ln\left(\frac{X}{\Lambda_{UV}}\right)\nonumber\\
&-&\frac{1}{2\pi(16\pi^{2})^{2}}\sum_{r=1,2}
\tilde{B}^{(3)}_{r}g_{r}^{4}\ln\left(\frac{\mu}{X}\right)+\cdots
\end{eqnarray}
We have neglected the one and two-loop terms coming from eq.\eqref{g1} since they are irrelevant for gluino mass.
$B^{(3)}_{r}$ modify due to the fact that when crossing the threshold $M$, we integrate out the messenger fields.
In terms of eq.\eqref{g2} we obtain the gluino mass,
\begin{eqnarray}{\label{g3}}
m_{\lambda_{3}}\simeq\sum_{r=1,2}\frac{\alpha_{3}}{4\pi}\frac{\alpha_{r}^{2}}{8\pi^{2}}\frac{F}{M}
\sim \frac{\alpha_{3}}{8\pi^{2}}\sum_{r=1,2}\alpha_{r}m_{r}
\end{eqnarray}
Note that the result eq.\eqref{g3} is valid with limit $F<<M^{2}$ (or $y\rightarrow 0 $) and
the sfermion and gagugino masses are input parameters at renormalization scale of $M$.

Now we analyze the constraints on Higgs messenger models,
some of which are also viable for the discussions in the next section.
\begin{itemize}
  \item Because of the negative contribution to sfermion masses coming from Yukawa interactions,
  the positivity of sfermion masses has to be imposed.
  Since the Yukawa couplings for first two generations are small compared with the third generation,
  we only take the third-generation scalar masses $\tilde{m}_{t}$ and $\tilde{m}_{b}$ into account.
  When $\tan\beta< 40$, $Y_b<< Y_{t}\simeq 1/\sin\beta$,
  which implies that the first coefficient term is extremely smaller than the second one in eq.\eqref{e2}.
  Thus we have in approximation,
  \begin{eqnarray}{\label{e6}}
  0<y<0.4
 \end{eqnarray}
 When $\tan\beta>>40$, one should analyze $\tilde{m}_{b}$ instead of $\tilde{m}_{t}$.
  Similar result can be found.

\item In order to induce electro-weak symmetry breaking via radiative correction, there must be a light
Higgs scalar. Without fine tuning, Higgs scalars masses are all of order of supersymmetry mediation scale $M$.
The corrections arising from renormalization group (RG) evaluation from $M$ to electroweak scale are negligible.
The scalar potential for Higgs scalar fields with $\mu=M$, $B\mu=F$ and soft masses given above can not
realize the EWSB as required.
\end{itemize}

This problem is solved in \cite{1012.2952} via introducing messengers charged under $SU(5)$ gauge group and
multiple spurion superfileds into the minimal Higgs messenger models.
In this paper, we will consider the possibilities of only extending Higgs sector of minimal Higgs messenger models.
As we will find, the models can successfully drive EWSB and have distinctive phenomenological features at LHC.

\section{Variant Higgs messenger models}
\subsection{Setup}
First, we define the Higgs sector as follows.
It contains four doublet superfields with representations under standard model gauge group,
\begin{eqnarray}{\label{e7}}
 H_{\mu}, R_{\mu}: (\mathbf{1},~\mathbf{2},~\frac{1}{2});~~~~~H_{d}, R_{d}: (\mathbf{1},~\mathbf{2},~-\frac{1}{2})
 \end{eqnarray}
The charges choices keep the models anomaly free.
Superfields $R_{\mu,d}$ directly couple to the SUSY-breaking sector as,
\begin{eqnarray}{\label{e8}}
W=XR_{\mu}R_{d}
\end{eqnarray}
The $H_{\mu,d}$ superfields that are responsible for EWSB are forbidden to either directly
couple to the SUSY-breaking sector or via the mixed couplings,
\begin{eqnarray}{\label{e9}}
W=X\left(H_{\mu}H_{d}+H_{\mu}R_{d}+H_{d}R_{\mu}\right),
\end{eqnarray}
Otherwise all of Higgs scalars masses will be of order of $M$,
and the problem of EWSB appears again as in the minimal Higgs sector models.
In this setup, the $R_{\mu,d}$ superfields will receive tree-level masses,
the $H_{\mu,d}$ scalars and Higgsinos will receive soft masses at two and one-loop respectively.
In other words, It is very natural to obtain such spectra,
\begin{eqnarray}{\label{e10}}
\tilde{m}_{Q}\sim m_{H_{\mu,d}}<< m_{R_{\mu,d}}\sim M.
\end{eqnarray}

In order to motivate the superpotential eq.\eqref{e8} and forbid superpotential eq.\eqref{e9},
we assume there exists a $Z_{n}$ parity, which could be a footprint of broken
global symmetry during dynamical supersymmetry breaking.
For example, we can assign superfields in Higgs sector $Z_{4}$-parity phases as follows
\footnote{More simple choices $Z_2$ and $Z_3$-parity are inconsistent with our analysis in these models.
More discussions about this aspect can be found below.},
\begin{eqnarray}{\label{z1}}
H_{\mu}\rightarrow e^{\frac{2\pi}{4}i}H_{\mu}, ~~~~H_{d}\rightarrow e^{\frac{2\pi i}{4}}H_{d}\nonumber\\
R_{d}\rightarrow e^{\pi i}R_{d}, ~~~~R_{\mu}\rightarrow e^{\pi i}R_{\mu}
\end{eqnarray}
Corresponding the phase factors for chiral matter superfields
under $Z_{4}$-parity transformation as $Q\rightarrow \exp{(\frac{s_{Q}\pi i}{4})}Q$
are determined to be,
\begin{eqnarray}{\label{z2}}
s_{Q}-s_{U}=2,~~~~s_{Q}-s_{D}=2,~~~~~s_{L}-s_{E}=2
\end{eqnarray}
This parity is expected to be broken furthermore at intermediate scale between $M$ and electroweak scale
\footnote{In this note, the $Z_4$ parity is broken around scale of SUSY-breaking mediation,
which is induced by some nozero VEVs during quantum generations of $\mu/B\mu$. }.

\subsection{Soft masses}

As shown in eq\eqref{e10},
$R_{\mu,d}$ fields are so heavy such that we can integrate them out from overview point of effective field theory.
The context of models are very similar to that of minimal supersymmetric standard model once
electroweak scale $\mu$ and $B\mu$
terms related to $H_{\mu,d}$ fields are correctly reproduced.
Before we proceed to discuss the generations of $\mu$ and $B\mu$ terms, we outline the
phenomenological features in these Higgs messenger models as follows.

The origin of sfermions masses is the same as in gauge mediation.
  The contributions coming from Yukawa interactions of $W_{MSSM}$ in eq.\eqref{e2} disappear.
  In this sense, there are three sum rules related to sfermion masses at scale $M$ according to
  the discussions of general gauge mediation \cite{GM2,GM3},
\begin{eqnarray}{\label{e11}}
\tilde{m}^{2}_{U}=4\tilde{m}^{2}_{D},~~~~ \tilde{m}^{2}_{E}=9\tilde{m}^{2}_{D},~~~~
\tilde{m}^{2}_{Q}+2\tilde{m}^{2}_{D}-\tilde{m}^{2}_{L}=0
\end{eqnarray}
 Furthermore, the Higgs scalars masses are simply related to the sfermion masses as,
\begin{eqnarray}{\label{e12}}
\tilde{m}^{2}_{H_{\mu}}=\tilde{m}^{2}_{H_{d}}=\tilde{m}^{2}_{L},
\end{eqnarray}
When RG evaluation to electroweak scale,
$\tilde{m}_{H_{\mu}}$ decrease to negative value more rapidly
in comparison with $\tilde{m}_{H_{d}}$ due to large top quark Yukawa interaction.

The first two gaugino masses are generated at one-loop, as given in eq.\eqref{e4}.
While the gluino mass for $SU(3)$ gauge group is generated at three-loop.
The spectra eq.\eqref{e2} (without Yukawa contribution), eq.\eqref{e4} and eq\eqref{g3} indicate that
the gluino are the next-to the lightest supersymmetric particles,
with mass of order $10^{-3}\tilde{m}_{Q}$ when one takes allowed upper bound value for
$M\sim 10^{9}$GeV in low-scale gauge mediation .
The ratio of gluino mass $m_{\lambda_{3}}$ over $m_{\lambda_{1,2}}$ increases
when RG evaluation from $M$ to eletroweak scale.
The effect arising from this enhancement induces that $m_{\lambda_{3}}\sim 10^{-2}\tilde{m}_{\lambda_{1,2}}$.
Explicitly, there is a typical spectra at electroweak scale with $\sqrt{F}\sim 10^{8}$GeV and $M\sim 10^{9}$GeV,
\begin{eqnarray}{\label{spectra}}
\tilde{m}_{Q}\sim \tilde{m}_{\lambda_{1,2}}\sim O(10-50)TeV;~~~~
\tilde{m}_{\lambda_{3}}\sim O(100)GeV.
\end{eqnarray}
This spectra implies that in Higgs messenger models we discuss here,  there
are only possible experiment signals associated with stable and light gluino at LHC \cite{1101.1645}.

The degeneracy of Higgs scalar masses at scale $M$ shown in eq.\eqref{e12}
and typical spectra eq.\eqref{spectra} suggest
that the parameter space where $\tan\beta$ near unity is favored due to the relation at electroweak scale\cite{9709356},
\begin{eqnarray}{\label{n}}\nonumber
m_{Z}^{2}=\frac{\mid \tilde{m}^{2}_{H_{\mu}}-\tilde{m}^{2}_{H_{d}}\mid}{\sqrt{1-\sin^{2}(2\beta)}}
-\tilde{m}^{2}_{H_{\mu}}-\tilde{m}^{2}_{H_{d}}-2\mu^{2}
\end{eqnarray}
In the next subsection, we will discuss a specific model that respects the $Z_4$ parity and
induces $\mu$ and $B\mu$ term as required.

Unless the initial values of soft masses at scale $M$ are very large ($>1$TeV),
the sum rules in eq.\eqref{e11} and  eq.\eqref{e12} approximately hold once RG evaluation is taken into account.
We also want to mention that, unlike the models proposed in \cite{1012.2952},
the flavor violations arising from the Yukawa interactions are of high order quantum effects,
which are negligible.
Thus, no stringent FCNC appears in these kind of models.

\subsection{$\mu/B\mu$ terms}
Now we discuss the generations of $\mu$ and $B\mu$ terms.
In order to drive EWSB as required,
$\mu$ and $B\mu$ should be induced at one and two-loop respectively.
A theoretic insight into this realization can be found in \cite{9603238,0711.4448},
in which the authors found that the effective action when integrating out the messengers should take
such a general form,
\begin{eqnarray}{\label{e14}}
\int d^{4}\theta ~H_{\mu}H_{d}\left[A(X)+B(X^{\dag})+D^{2}C(X,X^{\dag})\right]+h.c.
 \end{eqnarray}
in order to induce one-loop $\mu$ and two-loop $B\mu$ terms.

In \cite{9603238}, the mechanism proposed relies on the last term in eq.\eqref{e14}.
Two MSSM singlets $S$ and $N$ are introduced into the models with superpotential,
\begin{eqnarray}{\label{e15}}
W=S\left(\lambda_{1}H_{\mu}H_{d}+\lambda_{2}N^{2}+\lambda_{3} R_{\mu}R_{d}-M_{N}^{2}\right)
 \end{eqnarray}
The $S$ superfield should not carry any discrete or global quantum numbers for a successful realization of
$u/B\mu$ terms. It directly lead to that operator $H_{\mu}H_d$ has the same $z_4$ transformation as $R_{\mu}R_{d}$,
which should be forbidden in our models.

In \cite{0711.4448}, the mechanism relies on the second term in eq.\eqref{e14}.
In comparison with fields in \cite{9603238}, another bi-fundamental messengers $T_{\mu}$ and $T_{d}$ are introduced in
the SUSY-breaking superpotential $W=X\left(R_{\mu}R_{d}+T_{\mu}T_{d}\right)$.
The tree-level superpotential is given by,
\begin{eqnarray}{\label{e16}}
W=S\left(\lambda_{1}H_{\mu}H_{d}+\lambda_{2}N^{2}-M_{N}^{2}\right)+\lambda_{3}N R_{\mu}T_{d}
 \end{eqnarray}
In both these two models, if we assign the superfields with parities
that permit the superpotential eq.\eqref{e15} or eq.\eqref{e16},
then there will inevitably exists term like $XH_{\mu}H_{d}$,
which respect all symmetries.
Thus, these two specific constructions are both inconsistent with the setup of Higgs messenger models.

However, there is indeed a specific construction in the second mechanism that can induces correct $\mu/B\mu$.
With only one singlet $S$ and tree-level superpotential \cite{0711.4448},
\begin{eqnarray}{\label{e17}}
W=\lambda_{1}SH_{\mu}H_{d}+\frac{1}{2}M_{2}S^{2}+\left(M_{1}+\lambda_{2}S\right)R_{\mu}T_{d}+X(R_{\mu}R_{d}+T_{\mu}T_{d})
 \end{eqnarray}
As noted in \cite{0711.4448},
$M_1$ is a dynamical scale comparable with $M_2$ and carries an $U(1)_R$ phase factor.
After integrating out the messengers $R_{\mu,d}$ and $T_{\mu,d}$,
we obtain the effective Kahler potential.
The one-loop $\mu$ term is induced once $S$ is integrated out through its equation of motion.
The $B\mu$ term, on the other hand, can only be induced via two-loop wave-function renormalization effects.
Finally, one gets,
 \begin{eqnarray}{\label{e18}}
\mu^{2}\sim B\mu\sim \mathcal{O}(\tilde{m}^{2}_{H})
 \end{eqnarray}
Superpotential eq\eqref{e17} can be realized via following $Z_4$-parity  rotation assignments,
\begin{eqnarray}{\label{e19}}
S\rightarrow e^{i\pi} S,~~~T_{\mu}\rightarrow e^{i2\pi}T_{\mu}, ~~~~T_{d}\rightarrow T_{d}
\end{eqnarray}
The $U(1)_R$ rotation of $M_1$ ensures that $Z_4$ parity can be retained for $M_{1}R_{\mu}T_{d}$ term in eq.\eqref{e17}
under $Z_4$ parity phases assignments given by eq.\eqref{e19}.

\section{Conclusions}
In this paper, phenomenologies of supersymmetric models
in which Higgs sector (or part of it ) serves as
messengers in low-scale gauge mediation are explored.
We find in this simple scenario there are rich phenomenological features unexpected.
Our main results include that there are three sum rules for scalar masses,
and gluino is the next-to the lightest supersymmetric particle.
In this sense this scenario is appealing phenomenologically.
Moreover, once an appropriate parity is imposed at or above messenger scale,
which is spontaneously broken between messenger and electroweak scale,
the EWSB can be successfully induced with one-loop $\mu$ and two-loop $B\mu$ generated.
Along this line, there could be other possibilities that the gauge groups of messenger sector
are not an $SU(5)$, but subgroups of it, which are worthy to be discussed.

\section*{$\mathbf{Acknowledgement}$}
This work is supported in part by the
Fundamental Research Funds for the Central Universities with project
number CDJRC10300002.

\end{document}